\documentclass{elsart}

\usepackage{natbib}

 \usepackage{epsfig}

\usepackage{amssymb}

\begin{document}

\begin{frontmatter}















\title{Hard X-ray colours of Neutron Star and Black Hole Low Mass 
X-ray Binaries with \emph{INTEGRAL}}





\author[1,2]{Simon E. Shaw}
\author[3]{Ada Paizis}
\author[2,4]{Nami Mowlavi}
\author[2,4]{St\'ephane Paltani}
\author[2,4]{Thierry J.-L. Courvoisier}
\author[5]{Arvind Parmar}
\address[1]{School of Physics and Astronomy, University of Southampton, SO17 1BJ, UK.}
\address[2]{\emph{INTEGRAL} Science Data Centre, Chemin d'\'Ecogia 16, CH-1290 Versoix, Switzerland.}
\address[3]{INAF-IASF Sezione di Milano, Via Bassini 15, 20133 Milano, 
Italy}
\address[4]{Geneva Observatory, University of Geneva, ch. des Maillettes 51
CH-1290 Sauverny, Switzerland.}
\address[5]{ESTEC, P.O. Box 299, NL--2200--AG Noordwijk, Netherlands}

\begin{abstract}

The X-ray spectra of Low Mass X-ray Binaries (LMXB) can change on short time-scales, making it difficult to follow their spectral characteristics in detail through model fitting. Colour-colour (C-C) diagrams are therefore often used as alternative, model independent, tools to study the spectral variability of these sources. The \emph{INTEGRAL} mission, with its high sensitivity, large field of view and good angular resolution, is well suited to study the hard X-ray properties of LMXBs.  In particular the ISGRI imager on board of \emph{INTEGRAL} allows the regular monitoring of the sources in the less frequently studied domain above 20~keV. In this proceeding, C-C diagrams have been made with data from the \emph{INTEGRAL} public archive; a search is made for systematic differences in the C-C diagrams between black hole candidates  (BH) and neutron stars (NS) in LMXBs using a moments analysis method.
\end{abstract}


\end{frontmatter}

\section{Introduction}
\label{sec:int}
A Low Mass X-ray Binary (LMXB) is a gravitationally bound system comprising an evolved star and a compact object, where the mass ratio of the two objects is close to unity.  It is very difficult to observationally identify if the compact object is a stellar mass black hole (BH) or a neutron star (NS), and this problem remains one of the currently unresolved questions in astrophysics.  In the X-ray regime, it is possible to confirm that an LMXB contains a NS by observing phenomena that imply the existence of a physical surface on the compact object, such as Type 1 X-ray bursts, pulsations or kHz QPOs; however, the non-detection of these behaviours does not mean that a BH is present.

A useful way to disentangle classes of sources is to 
divide the spectrum in several broad energy bands and see how the `colour' (i.e. the ratio of 
the count rate in two different energy bands) correlates with another 
colour. These are 
the so-called colour-colour diagrams (C-C) and they enable a synthetic view of different 
sources, studying them as a class, away from single source peculiarities.
C-C diagrams are regularly used to investigate source properties and up to now 
they have mainly been made in the soft X-ray domain, below 20\,keV.
Indeed, it is thanks to such diagrams that Hasinger \& van der Klis (1989), using \textit{EXOSAT} data, discovered that bright NS LMXBs seem to divide themselves in two main classes, the so-called 
`Z' and `Atoll' sources (according to the pattern in their C-C plots). 
C-C diagrams are not a mere pattern-plotting but can pinpoint important physical properties. Indeed, Z and Atoll 
sources  never `jump' from one part of 
the diagram to the other and their position on the pattern is  likely linked to a 
\emph{physical} quantity that evolves in a continuous way, most likely the accretion rate (Hasinger et al., 1990). Furthermore, the source classification based on 
C-C patterns turned out to have physical grounds: Atoll sources have weaker
magnetic fields (about $10^6$ to $10^7$\,G versus 10$^8$--10$^9$\,G of
Z sources),  are generally fainter ($0.01$--$0.3 L_{\rm{Edd}}$
versus $\sim L_{\rm{Edd}}$), can exhibit harder spectra, trace out their C-C pattern on longer time
scales and have a different correlated
timing behaviour. Namely, different C-C patterns underline different physical conditions\footnote{
Muno et al. 2002 and Gierli\'nski \& Done 2002 observed that some Atoll sources, if
observed long enough, do exhibit a Z shape in the CC as
well. However, the above differences stated in the text, still remain.}\\
Similarly, Done and Gierli\'nski (2003)  built C-C plots for BH and NS LMXBs and HMXBs based 
on a large \textit{RXTE} archive (3--16\,keV). 
They show that portions of the C-C diagram are {\em never\/} covered
by NS systems, possibly due to their boundary layer emission.\\

The aim of this work is to search for similar effects in the less studied
 hard X-ray domain, above 20\,keV, using the \textit{INTEGRAL} data base in newly defined, harder, 
 energy bands.

\begin{figure}[pth]
\begin{center}
\includegraphics[scale=0.8]{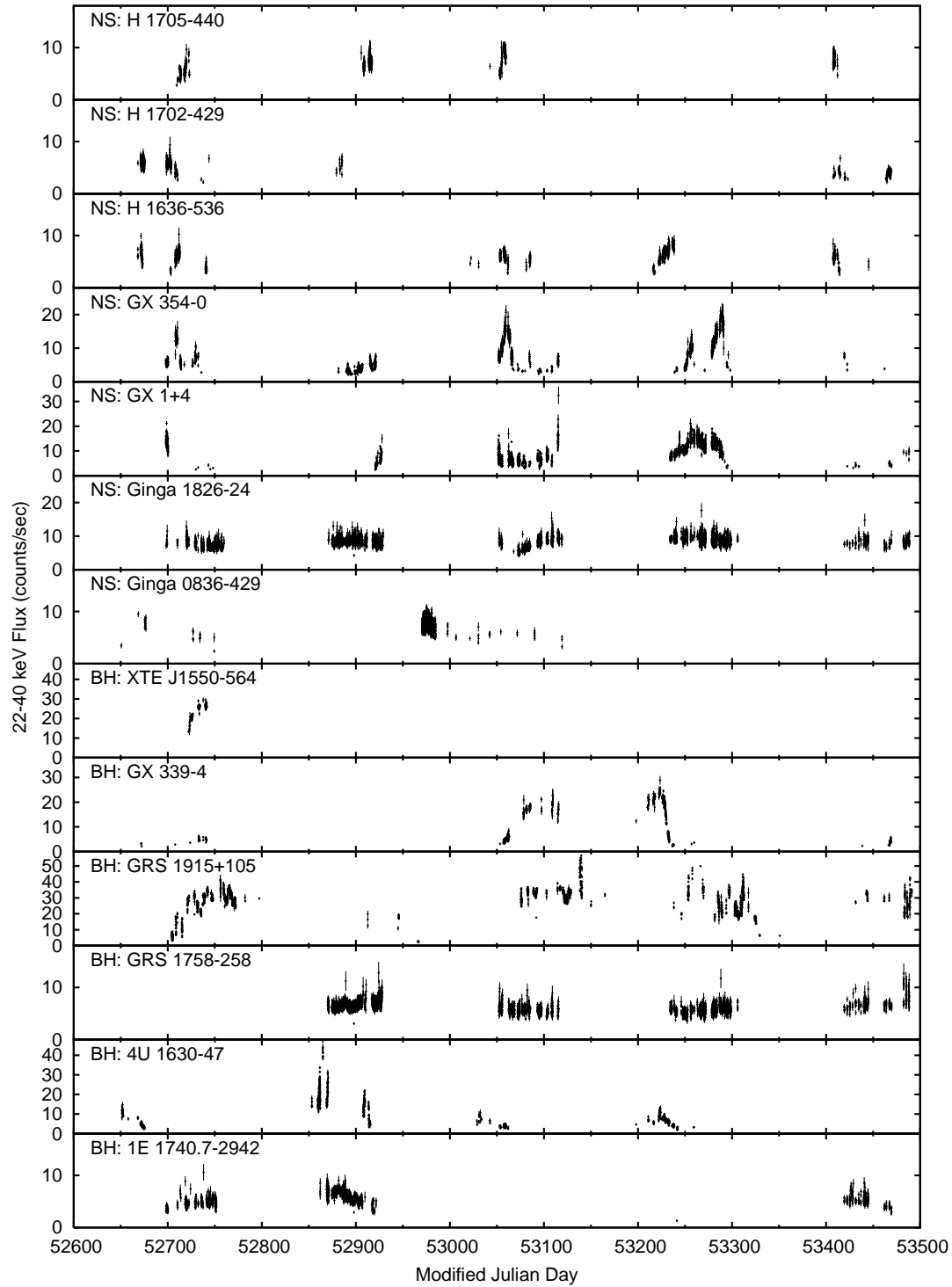}
\caption{Hard X-ray light curves of BH and NS LMXBs, taken from public data of the \emph{INTEGRAL} Source Results (\emph{http://isdc.unige.ch/index.cgi?Data+sources}).}\label{fig:lc} 
\end{center}
\end{figure}

For this study, data were taken from the \emph{INTEGRAL} Source Results (ISR) archive, which provides the hard X-ray fluxes measured in 30--60~minute time bins in several energy bands of all sources detected by either the ISGRI or JEM-X instruments.  This dataset currently contains 174 sources detected in the first 2 years of the mission and is available to the public \footnote{See \emph{http://isdc.unige.ch/index.cgi?Data+sources}}.  Some example 22-40~keV light curves are shown in Fig.~\ref{fig:lc}.  The \emph{INTEGRAL} satellite is a hard X-ray/soft $\gamma$-ray observatory that was launched by ESA on 17 October 2002 \citep{2003A&A...411L...1W}.  It was designed with several instruments to offer fine spectroscopy and imaging in the 3~keV--10~MeV range.  Of particular interest is ISGRI, which provides hard X-ray imaging in the 20--500 keV range in a large (30 degree) field of view \citep{2003A&A...411L.131U,2003A&A...411L.141L}.  The large field of view makes it ideal for studying transient events and detecting serendipitous objects. In the 20--60 keV range the ISGRI detector has the following imaging performance: angular resolution $\sim$~12$^{\prime}$; point source location accuracy $3^{\prime}$ for a $5\sigma$ sigma source; sensitivity $\sim$~18~mCrab for a 5$\sigma$ source seen in an 1.8~ksec observation \citep[e.g.][]{erik}.  Data from the JEM-X soft X-ray monitor \citep[3--30~keV;][]{2003A&A...411L.231L} were not considered, since the field of view is approximately 5 times smaller, and hence the availability of simultaneous data is reduced by the same factor.

\section{Hard X-Ray Colours}
\label{sec:cols}
Hard X-ray colours were computed from IBIS/ISGRI data, using the following expressions:

\begin{equation}
C_{s}  =  \log_{10}\left(\frac{F_{26-30}}{F_{22-26}}\right)   \textrm{~~~~and~~~~~}  C_{h}  = \log_{10}\left(\frac{F_{40-50}}{F_{30-40}}\right)
 \end{equation}

where $F_{X-Y}$ is the ISGRI count rate detected in a single pointing or Science Window (ScW; a stable pointing of roughly 30 -- 60~minutes, depending on the observing mode) in the $X-Y$ keV energy range.  Both $C_{s}$ and $C_{h}$ are only calculated if fluxes are available in all four bands; since most X-ray sources have steep spectra, the detection of an object above 40~keV is the most important factor in selecting the data.   Each colour is based on the ratio of two equal width energy bands, which means that generally $C_{s}$ and $C_{h} < 0$.

The energy range for this study is driven by the available data;  however, note that this approach may be particularly sensitive to spectral changes around 30 keV, such as a break or cut-off.  For each LMXB source, we select the observations where the source lies within 12$^{\circ}$ of the center of the field of view to avoid the effects of changing signal-to-noise towards the edge of the ISGRI field of view.  This leads to a sample of 14 LMXBs, with 6 BH candidates and 8 NS.

\begin{figure}[ph]
\begin{center}
\includegraphics[width=13cm]{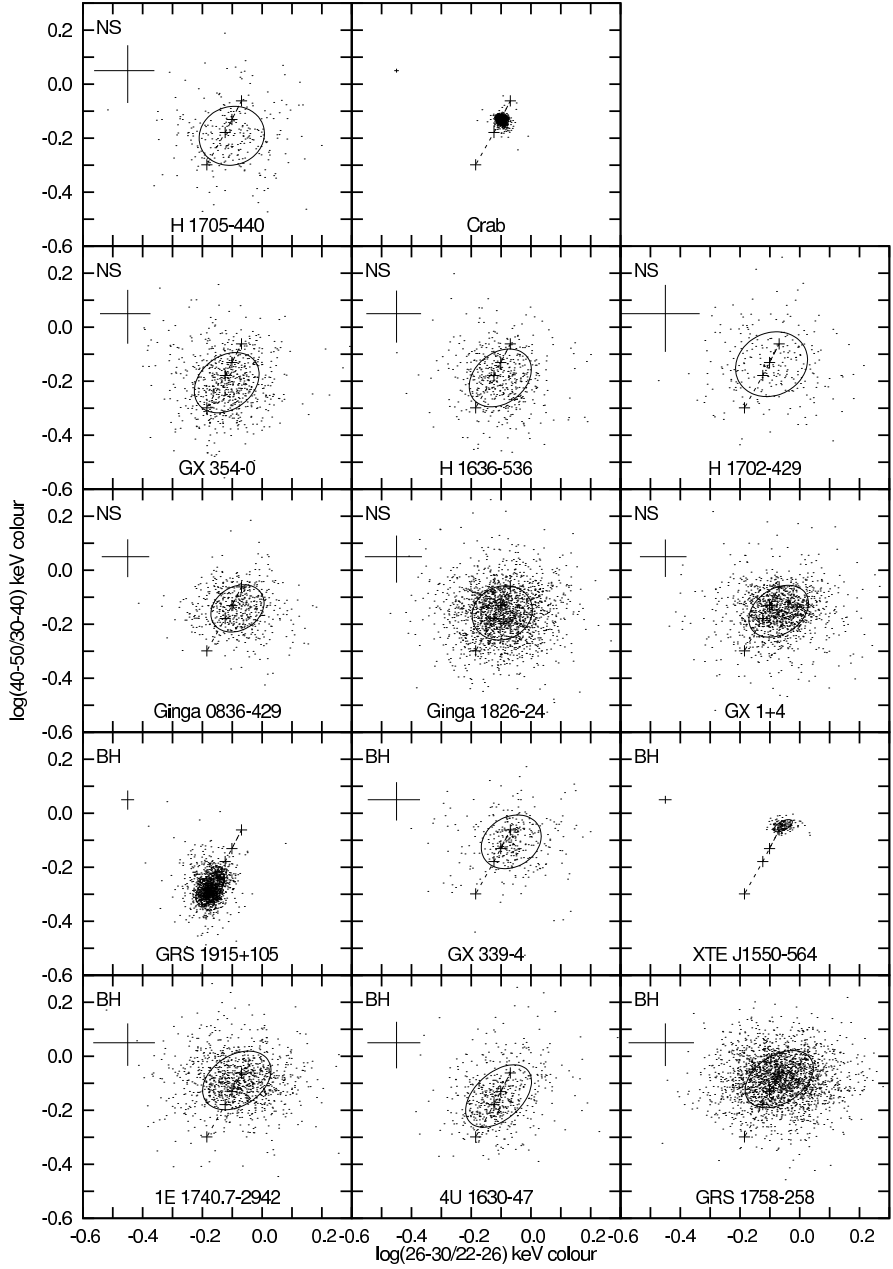}
\caption{Hard X-ray log colour colour plots of Neutron Star and Black Hole LMXBs, made with public data from the \emph{INTEGRAL} Source Results.  The name and type of object is indicated on each plot and the C-C plot for the Crab is shown in the top centre panel for comparison.  Each point represents the data from a single ScW pointing with the cross in the upper-left of each plot indicating the average error on a single point; systematic uncertainties are not included, but are thought to be similar in size to the cross shown for the Crab.  The dashed line is $P$, which represents where a source would lie if it had a simple power law spectrum.  The crosses on $P$ mark the index of the simulated power-law; from top-right to bottom-left,  $\Gamma = $1.5, 2.1 (Crab-like), 2.5 and 3.5.  The ellipses represent a parameterisation of the shape of the scatter plot and are described in detail in section~\ref{sec:moms}.}\label{fig:cc} 
\end{center}
\end{figure}

The measurement errors on the count rates in each energy band are calculated by the OSA analysis software as it extracts the flux for each source from a fitted sky position and are dependent on the detection significance (or brightness).  The uncertainty on  $C_{s}$ and $C_{h}$ can therefore be calculated from the square root of the quadratic sum of the fractional errors on each count rate.  Note that the bands used for $C_{h}$ are approximately twice as large as for $C_{s}$, this helps to compensate for the effects of increased error due to reduced counts at higher energies due to the steep spectra.  The average uncertainty on both $C_{s}$ and $C_{h}$ are approximately the same for each source, as shown in Fig.~\ref{fig:cc}.  However, there is a dependence of the uncertainties on the energy spectrum of each source and the sensitivity of the ISGRI instrument; sources with relatively hard spectra, such as XTE~J1550$-$564, have smaller uncertainty in $C_{h}$ whereas the opposite is true for soft sources, such as GRS~1915$+$105.  Systematic errors are not given by the OSA analysis, but are known to be present at the level of $\sim$~2\% in each band (Walter, 2006), implying $\sim$~3\% on  $C_{s}$ and $C_{h}$; for comparison, the average uncertainty indicated by the cross for the C-C plot of the Crab in Fig.~\ref{fig:cc} is $\sim$~1.5\% in both $C_{s}$ and $C_{h}$.

The distribution of $C_{s}$ and $C_{h}$ for a given object results in an approximately elliptical scatter plot, as shown in Fig.~\ref{fig:cc}.  All points are plotted wherever it is possible to calculate both $C_{s}$ and $C_{h}$.  The line in Fig.~\ref{fig:cc} is the effective power-law line, $P$, which is the result of simulating the value of $C_{s}$ and $C_{h}$ for sources with a simple power-law spectrum of differing photon indexes, $\Gamma =$ 1.5, 2.1 (Crab-like), 2.5 and 3.5.  The XSPEC software package was then used to create a model spectrum of the given $\Gamma$ and estimate the counts in each energy bin by using the ISGRI calibration files (RMF v16 and ARF v14).  The average count rates for the Crab in the ISR were compared against the simulations for $\Gamma = 2.1$ and found to agree within $\sim$~10\%.  The simulated count rates for each $\Gamma$ were then scaled by ratio of the simulated and average Crab count rates in the same energy bands.  By comparing the C-C plot of each source with $P$ it is possible to make a simple estimation of the average X-ray spectrum of the source: where a point lies on $P$, a good estimation of the spectrum will be a simple power-law with $\Gamma$ as indicated; those objects with large scatter in their C-C plots should have more complex and varying spectra.

\section{Moments analysis}
\label{sec:moms}
The elliptical shape of the C-C plots shown in Fig.~\ref{fig:cc} can be parameterised by measuring the moments that would be induced at its origin when considering it as a body made up of $i$ elements, of density $\rho_{i}$, each positioned at coordinates $(x_{i},y_{i})$.  In general

\begin{equation}
\langle x^{r}\rangle = \frac{\sum_{i}\rho_{i}x_{i}^{r}}{\Omega}
\label{eq:mom}
\end{equation}

The value $\Omega = \sum_{i}\rho_{i}$ contains no positional information, and is hence often referred to as the \emph{zeroth moment}.  The first and second order moments give the variances and covariance in $x$ and $y$, which in turn allow the calculation of parameters such as the length, width, orientation, etc. of an ellipse which describes the data.  

For each LMXB, the C-C plot shown in Fig.~\ref{fig:cc} is divided by a fine grid of size 0.05 colour units.  This creates a number of pixels, at coordinates $(C_{s_{i}},C_{h_{i}})$, containing $N_{i}$ data points, which allows the calculation of Eqn.~\ref{eq:mom}.  

\section{Results and Discussion}
 Several plots comparing simple measurements of the properties of the ellipses for NS and BH LMXB systems are shown in the following Figs.~\ref{fig:b}--\ref{fig:n}.  In all cases the uncertainty on the parameters has been estimated by calculating 1$\sigma$ confidence limits from the distributions returned by 2000 classical `bootstrap' trials; each sample contains the same number of points as the C-C data and hence contains some points from the original set more than once; more details of this method can be found in, for example, \citet{bootstrap}.

In Fig.~\ref{fig:b}, a plot of the centroids (first order moments) of the plots is shown, which represent the average behaviour of each object, along with $P$.  It can be seen that all the sources have both  $\langle C_{s}\rangle$ and $\langle C_{h}\rangle < 0$ and that they are strongly correlated, with a distribution that appears to be parallel to, and in general below, $P$.   With the exception of GRS~1915$+$105, which has a very soft spectrum, BH spectra seem to be harder than NS.  The NS objects considered here occupy a smaller range of indicated power-law than the BH and most are centred some distance from $P$; this suggests that a single power-law is not a good model for NS spectra, but it may be appropriate for at least some of the BH. All the NS are centred below $P$, which implies that the average spectrum that describes them would have a relatively harder $C_{s}$ and a softer $C_{h}$ when compared with a power-law; this phenomena would arise if $\Gamma_{> 30keV} > \Gamma_{< 30keV}$, e.g. if  there was a break in the average spectrum near 30~keV.

\begin{figure}[t]
\begin{center}
\includegraphics[scale=0.5]{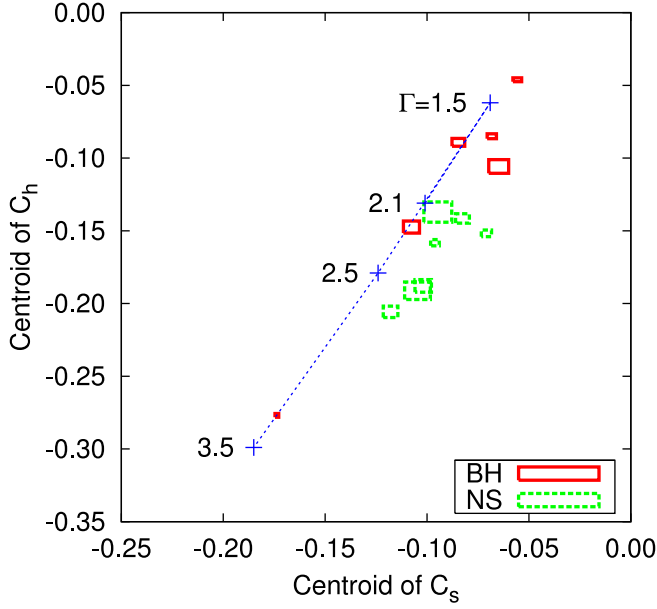}
\caption{The centroids of the hard X-ray C-C plots for NS and BH LMXBs, shown along with $P$ (dashed line - see also Fig.~\ref{fig:cc}); the sizes of the boxes are indicative of the 1$\sigma$ variation of the distribution of the centroids returned by a classical bootstrap analysis.  Crosses show the values of $\Gamma$ for the simulated power-laws.}\label{fig:b}
\end{center}
\end{figure} 


The rotation of the ellipses' major-axis, measured clockwise from the $C_{h}$ axis, is shown in Fig.~\ref{fig:f} and is expressed as the tangent of the rotation angle $\tan(r)$.  Because, in Fig.~\ref{fig:b}, all of the objects appear close to and are generally distributed parallel to $P$ and the projection of $P$ is in the direction of $(\langle C_{s}\rangle, \langle C_{h}\rangle) = 0$, a rough estimate of 'softness' of spectrum can be given by considering the Pythagorean distance, $d$, of the centroids from the origin of the C-C plots, with softer spectra appearing at larger $d$. 

All of the objects considered here have $\tan(r) >0$, which shows that in general a softening in $C_{s}$ is accompanied by a softening in $C_{h}$ and vice-versa. If the spectral behaviour of an object could be described solely by a variation in power-law index, all the individual measurements for that object would lie along $P$, i.e. $\tan(r) = 0.49$.  All of the objects here have  $\tan(r) > 0.6$ which suggests that even for those BH objects with centroids that lie on $P$ and whose average behaviour could be described by a power law, the general trend of spectral variability from ScW to ScW is more complicated.  Fig.~\ref{fig:f} also shows that there is a relationship between $d$ and $\tan(r)$; as the overall softness of the source (approximated by $d$) increases, the variation in $C_{h}$ also increases relative to $C_{s}$ leading to a rotation of the ellipse towards the  $C_{h}$ axis.  All of the objects seem to follow this trend, although the NS seem to occupy a smaller area of the plot, clustered around $\tan(r) \sim 1$, when compared to the BH.

It would also appear that the correlation between $C_{s}$ and $C_{h}$ is less strong for the NS systems.  Fig.~\ref{fig:m} shows  $\tan(r)$ plotted against the eccentricity, $\epsilon = \surd (1 - (b^{2}/a^{2}))$ where $a$ and $b$ are the lengths of the semi-major and semi-minor axes of the ellipse.  There is a large range of variation in $\epsilon$ in both types of source, but BH tend to have higher values than NS, with all NS having $\epsilon < 0.5$.  The NS have a tendency to have $\tan(r) \sim 1$, which means that on average a change in $C_{s}$ is accompanined by a similar fractional change in $C_{h}$.  However, it should be noted that as the ellipses become more circular, and $a \sim b$, their rotations tend towards 1, even though for a circle the 'rotation' is meaningless.

\begin{figure}[t]
\begin{center}
\includegraphics[scale=0.5]{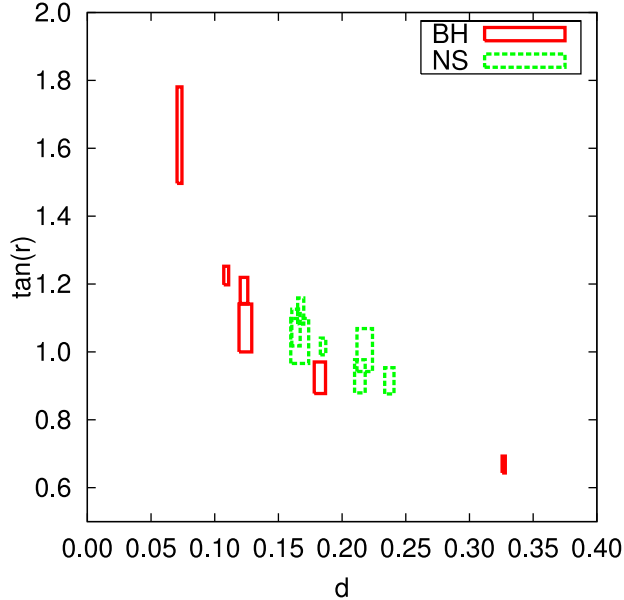}
\caption{Rotation of ellipse, $\tan(r)$, and distance of centroid from $(C_{s},C_{h}) = (0,0)$, $d$, of NS and BH LMXB hard X-ray C-C plots.  The sizes of the boxes are indicative of the 1$\sigma$ variation of the distribution of the parameters returned by a classical bootstrap analysis.}\label{fig:f}
\end{center}
\end{figure} 

\begin{figure}[t]
\begin{center}
\includegraphics[scale=0.5]{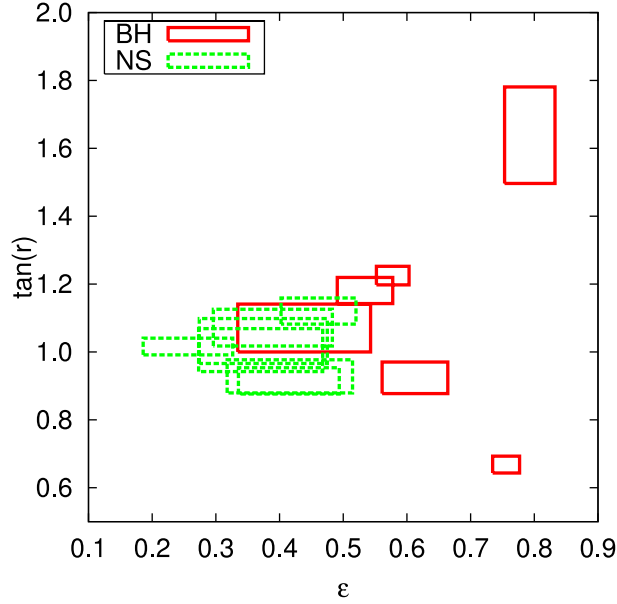}
\caption{Rotation, $\tan(r)$, and eccentricity, $\epsilon$, of NS and BH LMXB hard X-ray C-C plot ellipses.  The sizes of the boxes are indicative of the 1$\sigma$ variation of the distribution of the parameters returned by a classical bootstrap analysis.}\label{fig:m}
\end{center}
\end{figure} 

\begin{figure}[t]
\begin{center}
\includegraphics[scale=0.5]{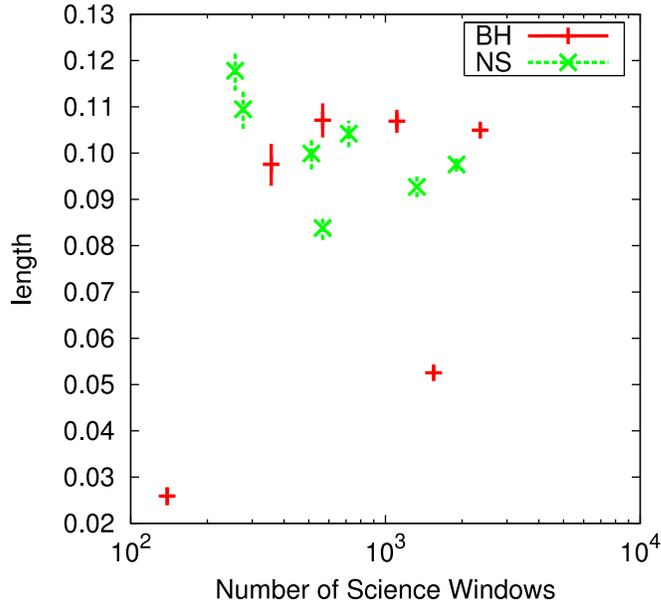}
\caption{Length of ellipse, $a$, in C-C plots for NS and BH LMXBs plotted against number of ScW, $N$, for each source.  The error bars on the length, $a$, are indicative of the 1$\sigma$ variation of the distribution of the parameters returned by a classical bootstrap analysis.}\label{fig:n}
\end{center}
\end{figure} 

\section{Conclusions and future work}
This  analysis was limited to those sources which were bright enough to have been seen up to the 40--50~keV band, in the \emph{INTEGRAL} public archive, in enough short $\sim$~1~hour pointings to allow a meaningful hard X-ray C-C plot to be constructed. This  results in a small sample of 6 BH and 7 NS, which may introduce a bias if the sources are only seen in particularly bright states.  The confidence in results of this analysis would clearly be improved with a larger sample of sources.  This could be achieved by integrating the individual points in the C-C plots over larger time scales in order to sample objects when in their weaker states.  There is a large range in exposure for the 14 selected LMXBs.  However, initial investigations suggest that parameters such as the length of the ellipse, $a$, are not correlated with exposure (Fig.~\ref{fig:n}).
   
To achieve a full understanding of this analysis, and any future analysis based on a larger sample of sources, it would advantageous to use the information contained within the measurement uncertainties on each individual flux band, rather than the simple bootstrap estimation method used here.  This could be achieved in a number of ways: by using the measurement uncertainties to calculate $\Delta~\langle x^{r}\rangle$ based on the binned C-C plots in Eq.~\ref{eq:mom}; by changing the analysis method to calculate moments based on individual points weighted by the uncertainty; or from the average of numerous numerical simulations of the parameters based on the randomised distribution of the count rate uncertainties.  As \emph{INTEGRAL} observations continue, the ISR public data base will grow and will potentially contain an increased number of sources;  the ISR is also under the process of being updated for the new release of the OSA6 software, which has an improved understanding of systematics on the count rates and improved off-axis response.  The best technique for considering the confidences in any parameter distribution differences between NS and BH will be investigated on the analysis of the new ISR data, but will not be discussed further in this proceeding. 

In conclusion, although the selection of data and energy bins for this analysis was driven by what was available from the public \emph{INTEGRAL} Source Results archive, it seems to have shown some potentially interesting differences between NS and BH LMXBs.  In short, the hard X-ray colours generated above and below 30~keV seem less well correlated for systems containing NS than those with BH; the average spectra of BH appear more `power law like' when compared with NS;  the centroids of NS C-C plots are grouped more tightly than BH, in a region below the line $P$, which would arise if the spectra of NS had a feature such as a break or cut-off around $\sim$~30~keV.

 While any separation can not currently be affirmed with a great deal of confidence, even the slight hope of being able to distinguish NS from BH simply based on their hard X-ray colours means that efforts must continue with a larger database and improved methods.





















\end{document}